\documentclass[prd, onecolumn, nofootinbib, floatfix, notitlepage]{revtex4-1}

\usepackage{amsmath}
\usepackage{graphicx}
\usepackage{dcolumn}
\usepackage{bm}
\usepackage{epsfig}
\usepackage{amssymb,latexsym,mathrsfs}
\usepackage{graphicx}
\usepackage{color}
\usepackage{xcolor}
\usepackage{hyperref}

\hypersetup{
    colorlinks=true,
    linkcolor=red,
    citecolor=blue,
} 

\newcommand{\be}{\begin{equation}}
\newcommand{\ee}{\end{equation}}
\newcommand{\bs}{\begin{split}} 
\newcommand{\bea}{\begin{eqnarray}}
\newcommand{\eea}{\end{eqnarray}}

\newcommand{\gm}{G_{\rm matter}} 
\newcommand{\gl}{G_{\rm light}} 
\newcommand{\geff}{G_{\rm eff}}

\newcommand{\al}{\alpha} 
\newcommand{\alm}{\alpha_M} 
\newcommand{\alb}{\alpha_B} 
\newcommand{\kap}{\kappa}
 
\newcommand{\eps}{\epsilon} 
 
\newcommand{\mpl}{M^2_{\rm pl}} 
\newcommand{\lamt}{\lambda^3} 
  
\newcommand{\dpds}{\dot\phi_{\rm dS}}  
\newcommand{\pds}{\phi_{\rm dS}}  
\newcommand{\dpo}{\dot\phi_1}  
\newcommand{\ddpds}{\ddot\phi_{\rm dS}}

\begin{document}

\title{Cosmic Acceleration with and without Limits} 

\author{Eric V. Linder${}^{1,2}$} 
\affiliation{
${}^1$Berkeley Center for Cosmological Physics \& Berkeley Lab, 
University of California, Berkeley, CA 94720, USA\\ 
${}^2$Energetic Cosmos Laboratory, Nazarbayev University, Astana, 
Kazakhstan 010000
} 

\begin{abstract}
A novel, interesting class of scalar-tensor gravity theories is those with a limit on the field motion,  
where the scalar field either goes to a constant acceleration 
or stops accelerating and goes to a constant velocity. 
We combine these with the ability to 
dynamically cancel a high energy cosmological constant, 
e.g.\ through the well tempered or 
self tuning approaches. One can successfully 
have a cosmic expansion history with a 
matter dominated epoch and late 
time acceleration despite a large cosmological constant, although the late time de Sitter limit may be unstable. 
Pole models, such as in a Dirac-Born-Infeld action, are 
of particular interest for a cosmic speed limit.  
\end{abstract} 

\date{\today} 

\maketitle


\section{Introduction} \label{sec:intro} 

The original cosmological constant problem -- how did 
the universe survive a high energy (possibly Planck scale) 
cosmological constant to give a universe that looks like 
ours -- is a fundamental issue in physics 
\cite{weinberg,cc2,cc3,cc4,cc5}. This is often glossed over in the quest 
to understand the late time, low energy cosmic acceleration. 
However, some approaches bring these together and look 
to resolve them through a dynamical cancellation by an 
evolving scalar field. 

This dynamical approach has been explored under the 
names of self tuning \cite{selftune1,selftune2,selftune3,st4,st5,st6,st7,st8,Khan:2022bxs,Appleby:2022bxp} and well tempered  
\cite{welltemper1,welltemper2,tempds,bflat,wtexp,wtber,tempsca} theories. These require higher order terms 
in the action involving the scalar field, and fit in 
well with the most general scalar-tensor theory leading 
to second order equations of motion, Horndeski gravity. 

Here we investigate a particular aspect of the 
dynamical approach, asking whether the scalar field 
evolution is unbounded, in field value, velocity, or 
acceleration. Speed limit theories have been studied 
in the context of inflation \cite{speedinflation,speed2,speed3}, 
without addressing the original cosmological constant 
problem, with one example being  Dirac-Born-Infeld 
theory. 

In Section~\ref{sec:shift} we introduce the equations 
of motion within shift symmetric Horndeski theory, and 
the conditions we seek: dynamical cancellation of the 
high energy cosmological constant, while preserving a 
matter dominated era, and asymptoting to a (low energy 
cosmological constant) de Sitter state. We then 
derive the conditions for a field value limit and a 
field speed limit in Section~\ref{sec:fieldnspeed}, and 
a field acceleration limit in Section~\ref{sec:accel}, 
achieving a standard 
cosmic expansion history despite a high energy 
cosmological constant. Section~\ref{sec:concl} 
summarizes and concludes, while 
Appendix~\ref{sec:propfn} investigates 
stability of the effective field theory de Sitter background.

\section{General Shift Symmetric Theory} \label{sec:shift} 

We begin within the most general scalar-tensor theory leading to second 
order equations of motion, Horndeski gravity, 
and examine the conditions that allow for dynamical 
cancellation of a high energy cosmological constant 
(i.e.\ removing the original cosmological constant problem). 
This self tuning imposes relations between the 
Horndeski action terms, and will set up the investigation 
of how the asymptotic future field motion -- either a field 
value limit $\phi\to0$, a speed limit $\dot\phi\to\,$const, 
or an acceleration limit $\ddot\phi\to\,$const   
-- further determines the behavior. 

The Horndeski action is 
\be 
S = \int d^{4}x \sqrt{-g} \,\Bigl[ G_4(\phi)\, R + K(\phi,X) -G_3(\phi,X)\Box\phi - \Lambda + {\cal L}_m[g_{\mu\nu}]\,\Bigr]\ , 
\label{eq:action} 
\ee 
where $K$, $G_3$, $G_4$ are the Horndeski functions 
(with $G_5=0$ and $G_4$ purely a function of the scalar 
field $\phi$, not its kinetic term 
$X=-(1/2)g^{\mu\nu}\phi_\mu\phi_\nu$, 
to keep gravitational waves propagating 
at the speed of light). The other quantities are the usual  
Ricci scalar $R$, high energy scale cosmological constant 
$\Lambda$, and matter Lagrangian ${\cal L}_m$. 

Shift symmetry is a valuable property that protects against 
quantum corrections, preserving a simple form for the theory. 
Under shift symmetry, the most general forms 
for the action functions are 
\bea 
K(\phi,X)&=&\kap(X)-\lamt\phi \label{eq:kshift}\\ 
G_3(\phi,X)&=&G_3(X) \label{eq:g3shift}\\ 
G_4(\phi)&=&(\mpl+M\phi)/2\ , \label{eq:g4shift}
\eea 
where $\mpl=1/(8\pi G_N)$ and $G_N$ is Newton's constant. 
Note the tadpole term $\lamt\phi$ in $K$, and the 
tadpole-like term 
$M\phi$ in $G_4$. There is no tadpole-like $b\phi$ in 
$G_3$ because in the action a term $\phi\Box\phi$ can be converted into a 
total derivative and a term linearly proportional to $X$, 
which can be absorbed into $K$ (cf.\ \cite{tempds}). 

The equations of motion in a Friedmann-Lema\^{\i}tre-Robertson-Walker 
(FLRW) universe become 
\bea 
3H^2(\mpl+M\phi)&=&\Lambda+2XK_X-\kappa+\lamt\phi+6H\dot\phi XG_{3X} 
-3HM\dot\phi+\rho_m  \label{eq:fried}\\ 
-2\dot H\,(\mpl+M\phi)&=&
\ddot\phi\,(M-2XG_{3X})-H\dot\phi\,(M-6XG_{3X})
+2XK_X+\rho_m+P_m \label{eq:fulldh}\\ 
0&=&\ddot\phi\,\left[K_X+2XK_{XX}+6H\dot\phi(G_{3X}+XG_{3XX})\right]\notag\\ 
&\qquad&+3H\dot\phi K_X+\lamt+6XG_{3X}(\dot H+3H^2)
-3M(\dot H+2H^2)\,.  \label{eq:fullddphi} 
\eea 
Here a subscript $X$ denotes a derivative with respect 
to $X$ and a dot is a derivative with respect to coordinate 
time $t$. The Hubble expansion rate $H=\dot a/a$ where 
$a$ is the scale factor, and $\rho_m$ and $P_m$ are the 
matter energy density and pressure. We define $g\equiv XG_{3X}$.

We will be interested in cosmological solutions that give 
a late time period of accelerated 
expansion, and in particular those that have a future de Sitter 
asymptote, $H\to h=\,$const. 
One main focus will be to characterize the asymptotic 
approach.

\section{Field Limit and Speed Limit} \label{sec:fieldnspeed} 

The Friedmann equation~\eqref{eq:fried} will play an extremely important 
role, as we need its terms to cancel the high energy 
cosmological constant $\Lambda$ and to leave a late time 
de Sitter state $H\to h$. Those terms will have to 
interact with each other -- giving a connection 
between the action functions $K$ and $G_3$ -- in order 
to achieve the cancellation dynamically and without 
fine tuning. 

We begin by considering the $\lamt\phi$ term. The 
field value can asymptotically go to either a 
constant $\pds$ (possibly zero) or to infinity. 
Unless both the field velocity $\dot\phi$ and 
acceleration $\ddot\phi$ go to zero, then we must 
have $\phi\to\infty$.

\subsection{Field Value Limit} \label{sec:value} 

If $\phi\to\,$const (possibly zero), then as 
mentioned $\dot\phi$ and hence $X$ must go to zero 
asymptotically. Thus all the terms involving $K$ and 
$G_3$ must go to either zero, or infinity, depending 
on whether they depend on negative or positive powers 
of $X$. If some go to infinity, they must cancel each 
other or they will overwhelm the de Sitter term 
$3\mpl h^2$ on the left hand side of the Friedmann 
equation. However, since additional  diverging factors 
enter the 
$\dot H$ and $\ddot\phi$ equations of motion, which 
have zero on their left hand sides in the de Sitter 
limit, such as a cancellation is not generally viable. 
That is, it appears we cannot achieve all three conditions of dynamical cancellation 
of $\Lambda$, an asymptotic de Sitter state, and 
solutions to the equations of motion at the same time. 

Thus the $K$ and $G_3$ terms must go to zero 
asymptotically. This is somewhat undesirable as 
then the action itself shows no sign of anything to cancel 
the cosmological constant: it is asymptotically just 
the Einstein-Hilbert action. Looking more closely at 
the Friedmann equation we see the remaining term to cancel $\Lambda$ is 
$(\lamt-3h^2 M)\phi=\,$const. That is, the cosmological 
constant $\Lambda$ is canceled by a constant of 
integration $\pds$, not really dynamically. This can 
be regarded as fine tuning. Therefore, field value 
limit theories to cancel the high energy cosmological 
constant and leave a low energy cosmological constant 
are not particularly satisfactory.

\subsection{Speed Limit} \label{sec:speed} 

We therefore proceed to speed limit theories. 
Here, terms proportional to $\phi$ will be 
growing with time. 
One possibility is to take $\lamt=3h^2M$, so 
as to preserve the de Sitter desiderata of 
$3\mpl h^2$ on the left hand side of the Friedmann 
equation. However, there is a hidden price: 
the gravitational coupling to the Ricci 
scalar, $G_4=(\mpl+M\phi)/2$, blows up as $\phi$ 
gets large (since $\dot\phi\to\,$const asymptotically). 
This will greatly disrupt the equations at the 
perturbative level, e.g.\ the growth of matter 
density perturbations. One might try to arrange that 
the blow up occurs in the future, beyond the reach 
of observations, but this requires fine tuning such 
as choosing initial conditions such that $M\phi_i$ 
is extremely small. 

This is a shame, as indeed a speed limit 
theory as simple as 
\bea 
K&=&-X-\lamt\phi\\ 
g&=&\frac{\lamt}{18h^2}+\frac{\dot\phi}{6h}\ , 
\eea 
will dynamically cancel a high energy cosmological 
constant, preserve radiation and matter domination, 
and lead to a late time cosmic acceleration and 
de Sitter state -- i.e.\ do everything we ask of it 
at the background level, while furthermore being 
ghost free and Laplace stable. 

Thus we will consider only theories 
with $M=0$. 
This will ensure that the left hand side, 
$3\mpl H^2$, of the Friedmann equation is 
constant for the asymptotic de Sitter state as we wish, 
but we then require cancellations among terms on 
the right hand side. 
That is, there must be a relation between the 
Horndeski terms $K$ and $G_3$, and the field $\phi$. 
At the same time, these must satisfy the equations 
of motion, i.e.\ the $\dot H$ and $\ddot\phi$ equations. 
This relation, or degeneracy, is the essence of the 
self tuning or well tempered mechanisms for dynamically 
canceling the high energy cosmological constant $\Lambda$. 

Here we take a novel branch of the self tuning class 
by restricting the asymptotic field motion to speed limit 
theories (and acceleration limit theories in 
Sec.~\ref{sec:accel}). 
For the approach to the 
de Sitter state we explore the field evolution by 
writing 
\bea 
\ddot\phi &\to& \ddpds+\dot F(t)\\ 
\dot\phi &\to& \ddpds t+F(t)+\dpo\\ 
\phi &\to& (1/2) \ddpds t^2 +\dpo t +\int dt\,F(t)\ , 
\eea 
where $\ddpds$ is a constant acceleration, 
$\dpo$ a constant speed, and $F(t)$ an 
evolution in speed to be determined by the 
equations of motion. 
We then  look for solutions with either a field speed 
limit or acceleration limit. 

For the speed limit case 
$\ddpds=0$, plus $F(t)$ must be subdominant to 
the constant term $\dpo$ at late times: thus the 
field velocity reaches a bound, $\dot\phi\to\dpo$. 
In the Friedmann equation the term $\lamt\phi$ will be 
growing as $t$. 
Therefore we have to arrange other terms in the 
Friedmann equation to negate the specific $\phi$ term 
growth. 
From Eq.~\eqref{eq:fried} we see this can be 
accomplished either by $\dot\phi$ or $\kappa(X)$. 
In fact, it cannot be done by $\dot\phi$ since 
if $\dot\phi\sim t^0$, then 
$\phi\sim t$  
grows faster. Thus we must choose $\kappa(X)$, and 
thence $g(X)$, to enforce the cancellation. 

To match $\phi\to\dpo t$ we need 
$\kap(X)\sim t$. Since $X=\dot\phi^2/2$, we can write 
this as 
\bea 
\dot\phi&=&\dpo+F(t)\\ 
\kap(X)&\sim&t\sim F^{-1}(\dot\phi-\dpo)\ , 
\eea 
where $F^{-1}$ indicates the inverse function 
(not $1/F$). Since $t$ gets large at late times 
while $\dot\phi-\dpo$ approaches zero, this 
indicates that speed limit theories require a 
pole function for $K_X$, e.g.\ 
\be  
K_X\sim(\dot\phi-\dpo)^{-p}\ . 
\ee  
There is nothing wrong with pole functions -- 
indeed Dirac-Born-Infeld kinetic terms are 
important in various areas of physics -- and 
they can work well cosmologically. 
One does have to be careful when solving the equations 
of motion numerically to write the equations in a 
form such that apparent divergences cancel. 

A simple speed limit model to cancel the cosmological 
constant and give an asymptotic de Sitter state is 
\bea 
K_X&=&\frac{c}{3h\dot\phi}-\frac{\lamt}{3h(\dot\phi-\dpds)}\\ 
g&=&-\frac{c\dpds}{18h^2\dot\phi}+\frac{\lamt\dot\phi}{18h^2(\dot\phi-\dpds)}\ , 
\eea 
where $c$ is a constant of dimension $\lamt$ and we 
have written $\dpo$ as $\dpds$ for clarity. Indeed, 
one can choose $c=\lamt$ and then 
\be 
K=-\frac{\lamt\dpds}{3h}\,\ln\left(\frac{\dot\phi}{\dpds}-1\right)-\lamt\phi\ . 
\ee 
In the expressions below, however, we keep $c$ general. 
While $K$ will involve a log (apparently, but see below), 
recall that in any theory where $g$ 
has a constant term then $G_3$ has a log also. 
Specifically, 
\bea 
K(\phi,\dot\phi)&=&\frac{c\dot\phi}{3h}-\frac{\lamt(\dot\phi-\dpds)}{3h}-\frac{\lamt\dpds}{3h}\,\ln\left(\frac{\dot\phi}{\dpds}-1\right)-\lamt\phi\\ 
G_3(\dot\phi)&=&\frac{c\dpds}{9h^2\dot\phi}+\frac{\lamt}{9h^2}\, 
\ln\left(\frac{\dot\phi-\dpds}{\dot\phi}\right)\ . 
\eea 

The scalar field approaches de Sitter with the behavior 
\bea 
\dot\phi&=&\dpds\left(1-e^{-3ht}\right)^{-1}\\ 
\ddot\phi&=&-3h\dpds\,e^{-3ht}\left(1-e^{-3ht}\right)^{-2}\\ 
\phi&=&-\frac{\dpds}{3h}\,\ln\left(\frac{\dot\phi-\dpds}{\dpds}\right)\ . 
\eea 
We now see that the $-\lamt\phi$ term on the de Sitter 
approach cancels the log term in $K$, leaving 
\be 
K(\phi,\dot\phi)\rightarrow\frac{c\dot\phi}{3h}-\frac{\lamt(\dot\phi-\dpds)}{3h}\ . 
\ee 

Thus, we have a theory with $M=0$, that has a speed limit 
$\dot\phi\to\dpds$, vanishing field acceleration 
$\ddot\phi\to -3h\dpds\,e^{-3ht}$, and in fact 
fulfills the well tempering degeneracy condition (rather 
than the trivial scalar field equation of 
self tuning). 
Note that terms in the equations of motion do not 
have divergences, but rather are of the form 
$\ddot\phi/(\dot\phi-\dpds)=-3h/(1-e^{-3ht})\to -3h$, 
$(H-h)/(\dot\phi-\dpds)\to -h/\dpds$, and 
$\dot H/(\dot\phi-\dpds)\to 3h^2/\dpds$. 

However, we are foiled once again at the perturbative 
level. The braiding property function \cite{bellsaw} 
is $\alb=2g\dot\phi/(H\mpl)$ and so as $g$ with its 
pole blows up (this is general, not dependent on the 
specific form adopted above), 
so does $\alb$. This not only gives 
rise to a Laplace instability, but during the evolution 
crossing $\alb=2$ the effective gravitational 
strength $\geff$ tends to diverge. Since $g$ nears 
its pole during the de Sitter approach, i.e.\ late time 
cosmic acceleration, it again would require fine tuning 
to push these disasters into the unobserved future. 
Property functions, stability, and gravitational strength are discussed further in Appendix~\ref{sec:propfn}.

\section{Acceleration Limit} \label{sec:accel} 

We now turn to the acceleration limit class 
of theories. To assess the possibilities, 
recall that 
an acceleration limit theory has $\ddpds={\rm const}\ne0$, 
with the function $\dot F(t)$ taken to die off at late 
times, i.e.\ $\dot F\sim t^{<0}$, 
and so the field acceleration reaches a bound, 
$\ddot\phi\to\ddpds$. The 
field speed then grows as $t$ at late times, and 
then the $\lamt\phi$ term in the Friedmann 
equation has a $t^2$ term. That must be offset by 
$2XK_X-\kap(X)+6h\dot\phi g$. The leading order term 
in $\dot\phi$ goes as $t$, so $2XK_X\sim t^2 K_X$, 
implying that a possible resolution is 
$K_X\sim\,$const (and hence 
$\kap(X)\sim X\sim t^2$). 
The other possible contribution is $\dot\phi g\sim tg$, 
requiring that $g\sim\dot\phi$. Let us write 
\bea 
K&=&bX+p(X)-\lamt\phi\\ 
g&=&g_1\dot\phi+g_0(X)\ . 
\eea 
Given the freedom in how $p(X)$, $g_0(X)$, $F(t)$, 
and $E(t)\equiv H(t)-h$ behave, there can be many classes 
of theories satisfying the constraining evolution 
equations near the de Sitter limit. 

To keep simplicity foremost, we take $p(X)=s\dot\phi$ 
and $g_0(X)=\,$const. One rationale is that 
fractional powers of $\dot\phi$ 
seem less motivated, and negative powers could give 
difficulties in the early universe where $\dot\phi$ 
may be small. The $p$ term also gives a $t^1$ term 
in the Friedmann equation and so can help cancel 
the additional $t^1$ term from $\lamt\phi$, 
while of course a 
constant in $K$ would just be absorbed in $\Lambda$. 
Basically, we are including all the terms that could contribute to canceling $\lamt\phi$ in the Friedmann equation. 
Finally, we are free to perform a field redefinition 
to normalize $|b|=1$; in fact we later find (see Appendix~\ref{sec:propfn}) that we 
need a negative sign, so $b=-1$, in order to be ghost free. 

Thus, our Ansatz is 
\bea 
K&=&-X+s\dot\phi-\lamt\phi \label{eq:ksimple}\\ 
g&=&g_0+g_1\dot\phi\ . \label{eq:gsimple} 
\eea 
We have seen that this has the potential to dynamically 
cancel $\Lambda$ in the Friedmann equation, so 
the next step is to examine the approach to the 
late time de Sitter limit. 

Evaluating the evolution equations term by 
term in powers of $t$, we find two cases. When 
$F(t)\sim t^{0<p<1}$, i.e.\ dominating over the 
constant term $\dpo$ in $\dot\phi$, then 
\bea   
H&\to&h+\frac{a}{t}+\frac{F}{t^2}\,\frac{a}{\lamt}\\ 
a&\equiv&\frac{1}{3}+\frac{6h^2g_0+hs}{\lamt}\ . 
\eea 
Furthermore, 
$h=1/(6g_1)$ and $\ddpds=-\lamt$. 
In fact, the Friedmann equation leads to 
$F(t)\sim \ln t$ being the only acceptable case 
of $F(t)\sim t^{0<p<1}$; i.e.\ no finite power 
$p$ works. 

If $F(t)\sim t^{<0}$, then 
we actually require $F(t)\sim t^{<-1}$ so 
that the Friedman equation term $\lamt\phi$ does 
not involve another (unmatched) term that grows 
with time. The approach to de Sitter is then 
\be    
H\to h+\frac{a}{t}+\frac{1}{t^2}\,\frac{a}{\lamt} 
\left(\dpo+6hg_0-\frac{\lamt}{3h}\right)\ ,  
\ee 
with $h=1/(6g_1)$, $\ddpds=-\lamt$, and $a$ as 
before, and the relation $s=-\lamt/(3h)$ must hold. 
This then gives $a=6h^2 g_0/\lamt$. 
However, if $g_0=0$ as well then the solution for the 
approach to de Sitter is not $(H-h)\sim t^{-1}$ 
but rather $(H-h)\sim e^{-3ht}$, i.e.\ an extremely 
rapid approach. 
For this case the field motion $\dot\phi\to\ddpds t+\dpo+\mathcal{O}(te^{-3ht})$. 
Note that 
if $s=0$, i.e.\ $K$ without the $p(X)=s\dot\phi$ term, the 
only acceleration limit case for these forms of 
$K$ and $G_3$ is $F(t)\sim \ln t$. 

One can obtain these results as well by recognizing 
that the $\ddot\phi$ equation is linear for this Ansatz 
and so its solution in quadrature is 
\bea 
(-1+6g_1H)\dot\phi&=&-6g_0H-a^{-3}\int dt\,a^3 (3Hs+\lamt)\notag\\ 
&=&-6g_0H-s-\lamt a^{-3}\int dt\,a^3\ , 
\eea 
together with the Friedmann equation 
\be 
3\mpl H^2=\Lambda+\lamt\phi+6Hg_0\dot\phi+\dot\phi^2\left(\frac{H}{h}-\frac{1}{2}\right)\ . 
\ee 

Table~\ref{tab:cases} summarizes the asymptotic 
behaviors for these cases. It seems remarkable 
that such simple forms for $K$ and $G_3$ can 
enable dynamical cancellation of the cosmological 
constant, and clear asymptotic approaches to de Sitter. 
Note that a subset of the acceleration limit with 
$F\sim\ln t$ case, fixing $s=0$ and $g_0=0$, was 
treated by \cite{Khan:2022bxs}; our results hold for general 
$s$ and $g_0$, which also reveal the two new 
acceleration limit behaviors.

\begin{table*} 
\centering 
\begin{tabular}{l|c|c|c} 
\hline 
\rule{0pt}{1.05\normalbaselineskip}Field motion limit &  $\ddot\phi$ & $\dot\phi$ & $H$ \\ 
\hline 
\rule{0pt}{1.05\normalbaselineskip}Speed limit  & $0+e^{-3ht}$ & $\dpds+e^{-3ht}$ & $h+e^{-3ht}$ \\ 
Acceleration ($F\sim\ln t$) &  $\ddpds+t^{-1}$ & $t+\ln t$ & $h+t^{-1}$ \\ 
Acceleration ($F\sim t^{<-1})$ &  $\ddpds+t^{<-2}$ & $t+\dpo$ & $h+t^{-1}$ \\ 
Acceleration ($s=-\lamt/(3h)$, $g_0=0$)\quad\, &  \quad $\ddpds+t e^{-3ht}$ \quad & \quad $t+\dpo+t e^{-3ht}$ \quad & \quad $h+e^{-3ht}$ \quad \\ 
\hline 
\end{tabular} \\  
\caption{Speed limit and acceleration limit cases 
of self tuning gravity, and the asymptotic behavior 
(leading order, and next to leading order, time dependence) 
of field motion and cosmic expansion approaching 
the late time de Sitter state.} 
\label{tab:cases} 
\end{table*}

What about the perturbative regime that caused 
us difficulties previously? Since $g\dot\phi$ grows 
with time, we will again have $\alb$ diverge, with 
the problems this causes. However, as this goes more 
slowly that the exponential behavior of the speed 
limit case, we have the possibility to sweep it to 
the future and keep the theory compatible with 
observations. The question of why the de Sitter background 
is eventually unstable, and what is the appropriate 
asymptotic background, becomes an issue that needs 
to be separately resolved (since it generally occurs 
when $\phi\gg M_{\rm Pl}$, other physics may need 
to enter).

\section{Conclusions} \label{sec:concl} 

Self tuning and well tempering can provide a path toward 
canceling a high energy cosmological constant, while 
preserving a matter/radiation dominated era, and then 
entering a late time cosmic acceleration with a de Sitter 
asymptote. An intriguing question is whether this dynamics 
can occur within limits: on either the field value, 
velocity, or acceleration. 

We show that no such ``cosmic acceleration with limits'' 
is perfect. At sufficiently late times the effective 
field theory breaks down with an instability in the 
de Sitter background. This might happen, e.g.\ for the 
acceleration limit class, late enough that it would 
not impact observations. Or it could be pointing to 
some indication that de Sitter space is not the 
correct asymptotic background or that higher order 
terms must enter the effective field theory. 

It is remarkable however that at the background 
level, quite simple models -- in either the speed 
limit or acceleration limit class -- can successfully 
provide a viable cosmic expansion despite the presence 
of a high energy cosmological constant. Models can 
be as simple as a minimal coupling to the Ricci scalar, 
(negative sign) canonical kinetic term with a tadpole, 
and $g=XG_{3X}\sim \dot\phi+\,$const, or a simple pole 
reminiscent of Dirac-Born-Infeld theory. 
The models presented are highly predictive, for 
example with just one free action parameter $\lambda$. 

We study in particular 
the asymptotic approach to a de Sitter state, 
deriving analytic behaviors for the field 
velocity $\dot\phi(t)$ and cosmic expansion $H(t)-h$. 
These results indicate that quite simple and 
tractable modified gravitational theories can 
have a rich and generally viable cosmology (at 
least in the expansion history) while treating 
the high energy cosmological constant problem. 
Future work could further explore pole models 
for viable examples, and consider the origin 
and fate of late time de Sitter instability.

\acknowledgments 

I am highly grateful to my well tempered collaborator 
Stephen Appleby for numerous insightful discussions 
on this topic. 
This work was supported in part by the Energetic Cosmos Laboratory and by the U.S.\ Department of Energy, Office of Science, Office of High Energy Physics, under contract no.\ DE-AC02-05CH11231.

\appendix

\section{Property Functions, Ghost Free, and Stability 
Conditions} \label{sec:propfn} 

At the background level, i.e.\ for the cosmic expansion, 
we have seen that we can successfully obtain a current, 
low energy epoch of 
cosmic acceleration -- despite the presence of a high 
energy cosmological constant -- and a de Sitter 
asymptote, i.e.\ the effective dark energy equation of 
state approaches $w=-1$. However we need to examine 
the perturbative level as well, i.e.\ small inhomogeneities 
away from the homogeneous FLRW background, 
to assess the impact 
on cosmic structure growth, not to mention testing 
the soundness of the theory through freedom from 
ghosts and gradient instability. 

It is convenient to investigate the perturbative 
quantities through the property functions \cite{bellsaw}. 
The Planck mass running, braiding (between the scalar 
kinetic and tensor sectors), and kineticity property 
functions are given for the 
shift symmetric class of 
Eqs.~\eqref{eq:kshift}--\eqref{eq:g4shift} with 
$M=0$ as  
\bea 
\al_M&=&0\\ 
\al_B&=&\frac{2\dot\phi g}{H\mpl}\quad \rightarrow\quad  \frac{2\dot\phi(g_0+g_1\dot\phi)}{H\mpl}\\ 
\al_K&=&\frac{2X(K_X+2XK_{XX})+12H\dot\phi Xg_X}{H^2\mpl}\quad \rightarrow\quad  
\frac{\dot\phi^2}{H^2\mpl}\,\left(6Hg_1-1\right)=\frac{\dot\phi^2}{H^2\mpl}\,\left(\frac{H}{h}-1\right)\ , 
\eea 
where recall $g\equiv XG_{3X}$. The right arrow 
specializes to the simple acceleration limit model 
of Eqs.~\eqref{eq:ksimple}, \eqref{eq:gsimple}, and 
we have imposed the requirement $g_1=1/(6h)$ in the last equation. 

At early times, all property functions are small, going 
as $1/H$. That is, general relativity is a good 
approximation in the early universe. At late times, 
$\al_K$ vanishes on shell, while $\al_B$ gets large 
as $\dot\phi g$ does. 

To ensure the theory is ghost free, it must satisfy 
the no ghost condition $\al_K+(3/2)\al_B^2\ge0$. 
Since $\al_K\ge0$ for the acceleration limit model 
used in Sec.~\ref{sec:accel}, this is always satisfied. 
Indeed, this was our motivation for choosing a negative 
coefficient $K\sim -X$. If one writes $K\sim \eps X$ 
then the equations of motion give $g_1=-\eps/(6h)$ and 
$\al_K\sim -\eps$. Thus we need $\eps=-1$ to be 
ghost free. 
One can also verify that the speed limit pole model 
used in Sec.~\ref{sec:speed} is ghost free. 

The Laplace stability 
condition delivering a nonnegative scalar sound 
speed squared is 
\be 
\left(1-\frac{\al_B}{2}\right)\,\al_B+\frac{(H\al_B)\,\dot{}}{H^2}-\frac{2\dot H}{H^2}-\frac{\rho_m+P_m}{H^2}\ge 0\ . 
\ee 
Now we see the problem with $\dot\phi g$, and hence 
$\alb$, getting large. The condition then becomes 
\be 
-\frac{\alb^2}{2} \ge 0\ , 
\ee 
which is clearly violated. Thus at a minimum 
the linear effective field theory behind the property 
functions breaks down at late times. 
For the acceleration limit theory, $\dot\phi$ gets 
large, while for the speed limit case, although $\dot\phi\to\,$const, $g$ gets 
large due to the pole. 
At early times, the last 
two terms in the Laplace condition will dominate, 
with the main contribution to their difference 
coming from 
the $\ddot\phi$ term in Eq.~\eqref{eq:fulldh}, 
leaving 
\be 
-2g\ddot\phi\approx -2g_0\ddot\phi \ge 0\ , 
\ee 
as the stability condition: i.e.\ $g_0\ddot\phi\le0$. 
At intermediate times one must check the condition 
numerically. 

Since $\al_M=0$, these models are in the class of No Run 
Gravity \cite{1903.02010}, albeit special ones 
that cancel a high energy cosmological constant. 
The effective gravitational coupling strengths are 
\cite{bellsaw,lmg} 
\be 
\gm=\gl=\frac{\alb+\alb'}{\alb(1-\alb/2)+\alb'}=
1+\frac{\alb^2}{\alb(2-\alb)+2\alb'}\ . \label{eq:glgen} 
\ee 
In the matter dominated era $\alb\ll1$ 
so $\gm=\gl\to 1$. In the de 
Sitter state, as expected from the Laplace 
condition, the strength of gravity blows up as 
$\alb$ increases too far, and then eventually 
approaches zero from below. 
(A theory with coupling $M\ne0$, 
and hence $\alm\ne0$, could avoid this 
but still tends to cause 
the gravitational strength to differ from general 
relativity too early; also see Appendix~C of 
\cite{tempds} 
for some further discussion). 
As long as initial 
conditions allow a matter dominated era, the 
values of $\gm$, $\gl$ should not be too extreme 
where we have data, i.e.\ the past, 
since the acceleration limit model grows $\dot\phi$ 
linearly at late times, but this should 
be numerically evaluated for any theory.


\end{document}